\documentstyle[preprint,aps,psfig]{revtex}
\begin{document}
%\draft
\title{Kink Arrays and Solitary Structures in Optically Biased Phase 
Transition}
\author{L.M. Pismen}
\address{Department of Chemical Engineering, Technion -- Israel Institute of Technology, 
32000 Haifa, Israel} 
\date{ }
\maketitle
\begin{abstract}
An interphase boundary may be immobilized due to nonlinear 
diffractional interactions in a feedback optical device. This effect reminds 
of the Turing  mechanism, with the  optical field playing the role of a 
diffusive inhibitor. Two examples of pattern formation are considered in 
detail: arrays of kinks in 1d, and solitary spots in 2d. In both cases, a large 
number of equilibrium solutions is possible due to the oscillatory 
character of diffractional interaction. 

\end{abstract}
\pacs{42.65.Pc,05.70.Fh,05.45+h}
\narrowtext
%==================================================

There is a growing interest in transverse effects in nonlinear optics that 
manifest themselves in spontaneous pattern formation in both active  
optical devices (lasers) and feedback systems (ring and 
Fabry -- Perot cavities) driven by coherent sources. 
Symmetry breaking phenomena in passive nonlinear optical systems
were studied in connection to self-focusing effects, stimulated Raman
and Brillouin scattering and optical bistability in feedback systems 
\cite{newmol,abfir,arec,bram,weiss}. A 
balance of non-linearity and diffraction that is needed for pattern 
formation arises through sequential operation of the two effects in an 
optically thin nonlinear medium and in the empty part of the cavity.  

Transverse optical patterns have attracted particular attention in view of 
a possibility to imitate a variety of other pattern formation processes in 
non-equilibrium systems that, typically, require far longer observation 
times. Optical experiments are, however, at disadvantage in that they have 
relatively small aspect ratios. The limiting factor here is the diffractional 
length that fixes the smallest scale of inhomogeneities of the intensity of 
the optical field.  At the same time, far steeper inhomogeneities, limited by 
a short diffusional scale, are possible in a nonlinear optical 
medium. While the optical field cannot {\em cause} such 
inhomogeneities, it may {\em stabilize} them if they arise spontaneously, 
for example, near a 
Maxwell point in either equilibrium or non-equilibrium phase transition. 

The situation 
we envisage reminds of a common mechanism of formation of Turing 
patterns \cite{seg,fife}: a nonlinear medium that can exist in two 
alternative states plays the role of a ``slowly diffusing activator'', while the 
role of the optical field is that of a ``rapidly diffusing inhibitor''. The 
analogy is, of course, not exact: nonlinear diffractive interaction in a 
feedback optical device is far more complex than diffusive interaction, and 
leads to a great number of alternative stable configurations.

Our basic model is an optical ring cavity containing a thin sample of
nonlinear medium with the thickness $l$ in the direction of beam
propagation negligible compared to the length $L$ of the optical path
in the empty part of the cavity. Under these conditions, diffraction of
the light beam inside the nonlinear medium can be neglected, and 
consecutive transformations of the signal during a round-trip in the cavity 
include diffraction in the empty part of the cavity, mixing with the pump 
beam, and point transformation in the nonlinear medium.

In the  paraxial approximation, the complex envelope amplitude
$E$ of the electric field in the empty part of the cavity obeys the
parabolic equation
\begin{equation}
iE_z= \nabla^2  E .
\label{basicdiff}  \end{equation}
Here the coordinate $z$ in the direction of propagation is scaled by the
length $L$, and the transverse coordinates, by the diffraction length
$\sqrt{L/k}$; $\nabla^2 $ denotes the transverse Laplacian. 

Since the material response time is typically much larger than the 
round-trip time in the cavity, we shall be interested in stationary solutions 
of Eq.~(\ref{basicdiff}) that give the field $U({\bf r} )$ in the 
nonlinear medium corresponding to an instantaneous distribution of the 
material variable $\chi({\bf r} )$. We define $U$ as the field obtained by 
mixing the attenuated cavity signal $E_f({\bf r})=E({\bf r},1)$ with the field 
$P$ of the pump beam. Then $U({\bf r})= P +\alpha E_f({\bf r})$ where 
$\alpha$ is the transmission coefficient. The field $E_i=E({\bf r},0)$ at the 
start of the next round trip is obtained by adding a phase shift $\beta\chi$ 
due to interaction with the medium.  Expressing the solution of 
Eq.~(\ref{basicdiff}) by a functional 
$E_f({\bf r}) =\Psi\left[E_i({\bf r})\right]$, and adding also a constant 
phase shift  $\Omega$ yields the functional equation for the 
quasistationary field $U({\bf r} )$: 
 \begin{equation}
U({\bf r}) =  P +  \alpha e^{i \Omega}\Psi
           \left[U({\bf r})e^{i \beta \chi({\bf r})}\right] .
\label{Kerr}    \end{equation}

We adopt a model of material dynamics allowing for a phase transition 
biased by the intensity of the electric field in the medium. The simplest 
model of this type can be written as an evolution equation for the material 
variable $\chi$ with a cubic nonlinearity, including a weak bias dependent 
on the intensity of the electric field: 
\begin{equation}
 \chi_t= \sigma^2\nabla^2  \chi + \chi -\chi^3 + w(|U|).
\label{material}  \end{equation}
In the following, we shall use the bias function
 $ w(|U|) =(|U|^2 - |U_0|^2 ) $. 
The variable $\chi$ can be interpreted as a phase field, in the spirit of 
common phenomenological phase-transition models \cite{cafi}. The material response 
time is taken as unity; $\sigma$ is the ratio of the 
diffusional and diffractional lengths. Typically, $\sigma \ll 1$, so that the interphase 
boundary is sharp. As usual, we assume a thin sample approximation; then 
Eq.(\ref{material}) retains only the transverse Laplacian $\nabla^2$.

Eq.~(\ref{material}) is a dissipative dynamic equation of the form $\chi_t = -{\delta\cal 
F}/\delta\chi$ derivable from a suitable expression for the free energy $\cal F$ (playing the 
role of a Lyapunov functional):
$${\cal F} = \int \left[\frac{1}{2}\sigma^2|\nabla \chi|^2 + 
   \frac{1}{4}(1 -\chi^2)^2 + \chi (|U|^2 - |U_0|^2) \right] dV.$$
The last term, responsible for the field-induced bias, is recognized as a standard correction 
to the free energy of a dielectric in the electric field when the dielectric permittivity is 
linearly dependent on $\chi$. The model is therefore applicable wherever two dielectric 
phases with different permittivities may coexist. Experimentally, this situation can be 
realized, for example, near an isotropic -- nematic transition. In this case, imposing a 
suitable orientation of the nematic phase can be used to enhance the difference between the 
refraction indices of the two phases. A more exotic possibility is a {\it 
non-equilibrium} phase transition between alternative steady states of a photochemical 
reaction. 

If $U$ is eliminated with the help of Eq.~(\ref{Kerr}), slow transverse 
dynamics is determined by the evolution equation (\ref{material}). The 
situation is somewhat similar to typical models of large aspect ratio 
dissipative pattern-forming systems where it is often possible to eliminate 
the ``vertical'' structure (in the direction of a flux sustaining the system in 
a non-equilibrium state). A substantial difference is that the effect of 
diffraction hidden in Eq.~(\ref{Kerr}) is non-local, which makes the 
analysis far more difficult. 

Once the diffractional problem is solved, Eq.~(\ref{material}) can be treated in a standard 
way \cite{fife,cafi}. At $\sigma \ll 1$, $\chi$ 
approaches one of alternative equilibrium values everywhere except 
boundary regions of $O(\sigma)$ thickness (kinks). The phase equilibrium 
at $|U|=|U_0|$ is biased by variable local field intensity, which sets the 
kinks into motion. The dynamics of a weakly bent kink is completely 
determined by its local geometry and the intensity of the biasing field 
$|U|$ at the kink position. 
The general equation of motion for a weakly curved kink is derived in a 
most transparent way with the help of multiscale expansion in a 
coordinate frame aligned with the kink \cite{pi94}. The kink 
propagates with the speed 
\begin{equation}
c = - \sigma \kappa \mp b w(|U|), 
\label{eqmot} \end{equation}
where $\kappa$ is the curvature of the kink, and the coefficient $b$ 
equals, for the cubic model in Eq.(\ref{material}), to $3/\sqrt{2}$. The 
upper sign in Eq.(\ref{eqmot}) applies to a kink with $\chi(\pm \infty)= 
\pm 1$, and the lower, to an antikink with the reverse orientation. 
Curvature effects may balance the biasing field at $\kappa \approx 
w/\sigma$. This balance can be achieved when a curvature radius is 
measurable on a long (diffractional) scale, provided the bias is weak. 

In the following, I shall restrict to the dissipative limit when the 
transmission coefficient is small; the pump field is assumed to be constant 
in space. Then Eq.~(\ref{Kerr}) yields, to the leading order in $\alpha$,
\begin{equation}
U({\bf r}) =  P \left(1 +  \alpha u({\bf r})  \right),  \;\;\; 
    u({\bf r})= e^{i \Omega}\Psi\left[e^{i \beta \chi({\bf r})}\right] .
\label{Kerr0}    \end{equation}
I shall consider first a one-dimensional model, and assume that the two 
phases coexist on an infinite $x$ axis, with an interphase boundary (kink) 
at $x=0$. In the zero order, the coexistence condition is $|P|=|U_0|.$ 
Assuming it holds, the profile of the material variable valid on the long 
(diffractional) scale is simply $\chi(x)=2H(x)-1$, where $H(x)$ is the 
unit step function: $H(x)=1$ at $x>0$, $H(x)=0$ at $x<0$. The 
jump is effected in a narrow $O(\sigma)$ interval that is negligible on 
the diffractional scale. Then Eq.~(\ref{basicdiff}) has to be solved with the 
initial condition 
\begin{equation}
E(x,0)=P \left[ e^{- i\beta} + 2i \sin \beta \, H(x) \right].
\label{incond}  \end{equation}
The solution is expressed through the error function of a complex 
argument. This gives the cavity field $u=u_1(x)$ in Eq.~(\ref{Kerr0}) 
\begin{equation}
u_1(x)  =  e^{i\Omega} \left[ \cos \beta + i \sin \beta \,\mbox{erf}
    \left( \mbox{$\frac{1}{2}$}\sqrt{i}x \right) \right].
 \label{vrr}    \end{equation}

Since Eq.~(\ref{Kerr}) is linear, solutions  corresponding to $n$
alternating kinks and antikinks are obtained by superposition:
\begin{equation}
u_n(x)  = \sum_{j=1}^n q_j u_1(x-x_j) ,\;\;\; q_j=(- 1)^{j-1}.
\label{nvrr}    \end{equation}
The corresponding bias of the field intensity is $w =2 \alpha|P|^2 
\mbox{Re} \,u_n(x)$. An $O(\alpha)$ bias of the electric field at the kink 
location $x=x_k$ causes it to migrate, according to Eq.~(\ref{eqmot}), with 
the speed 
\begin{equation}
c=-3\sqrt{2}q_k\alpha|P|^2 \mbox{Re} \,u_n(x_k)
\label{eqm0}  \end{equation}

Incorporating various constants into the time scale brings the equation of 
motion of a system of kinks and antikinks to the form
\begin{equation}
\frac{dx_k}{dt}  = -\sum_{j\neq k} q_j q_k \psi (x_k-x_j) ,\;\;\;  
     \psi(x) =  \mbox{Re} \, u_1(x)  .
\label{manym}    \end{equation}
This equation can be written in the gradient form
\begin{eqnarray}
\frac{dx_k}{dt}  &=& - \frac{\partial V}{\partial x_k}, \;\;\;
 V  = \sum_{j, k} q_j q_k \Phi (x_k-x_j) , 
\label{manyV}    \\
 \mbox{where} & &  
    \Phi(x) =  \int_0^x\psi(\xi) d\xi \nonumber \\
    &=&   x \psi(x) + \frac{2}{\sqrt{\pi}}
       \sin\beta \sin\left( \frac{x^2+\pi}{4} - \Omega\right).
\label{manyP}    \end{eqnarray}

The function $\psi(x)$ is a combination of Fresnel integrals dependent 
on the phase angles $\beta$ and $\Omega$. Both real and imaginary parts 
of the function erf$(\frac{1}{2}\sqrt{i}x)$  oscillate with increasing 
frequency and decreasing amplitude at $|x| \to \infty$ around the 
asymptotic real value erf$(\infty)=1$. These oscillations lead to a high 
complexity of solutions. Unlike long-range diffusive interaction in Turing 
patterns, the sign of diffractional interaction between kinks alternates with 
varying separation, and the potential Eq.~(\ref{manyP}) may have a 
great number of minima (see Fig.~1). 

We shall be mostly interested in the case when a kink and an antikink 
repel each other at short distances, which would lead to nucleation of 
kink-antikink pairs. This requires the derivative of the biasing field 
$\psi'(0)$ to be positive. Assuming $0<\beta<\pi$, this requires 
$-\frac{1}{4}\pi < \Omega < \frac{3}{4}\pi$. The repelling 
action may prevail, of course, on the diffractional scale only. Short-range 
attractive interaction takes over at $O(\sigma)$ distances. The balance of 
attraction and repulsion determines the critical size of a nucleus. In one 
dimension, short-range interaction falls off exponentially \cite{ohta}, and 
the size of the critical nucleus is $a_{cr} = O(\sigma \ln \alpha)$ is very 
small. 

Equilibrium separations of a kink/antikink pair, or, equivalently, the size 
of a solitary structure formed by the bound pair, are defined as zeroes of 
the biasing field $\psi(x)$. In a special case $\Omega+\beta= \pm \pi/2$, 
the asymptotic value of $\psi(x)$ at $x\to \infty$ vanishes. In this case, 
the equation $\psi(x)=0$ has an infinite number of solutions.
Asymptotically at large separations, 
\begin{equation}
\psi(x) = \cos(\beta + \Omega) - \frac{2}{\sqrt{\pi}x} \sin \beta
  \sin \left( \frac{x^2 +\pi}{4}-\Omega \right).
\label{asymp}  \end{equation}
 By this formula, equilibrium separation distances are $x_k = \sqrt{(4k-
1)\pi+\Omega}$, $n=1,2, \ldots$. Alternating points correspond to stable 
equilibria. The asymptotic values are very close to ``exact'' solutions, 
except for $x_1$; e.g. at  $\Omega=0, \beta=\pi/2$ the exact value is 
$x_1=3.17$, as compared to $\sqrt{3\pi}=3.07$.

Although the average interaction strength decays with separation not too 
fast ($\propto x^{-1}$), equilibrium positions of neighboring kinks in a 
multikink array are shifted only by few percentage points from positions 
computed taking into account pair interactions only. The minimal
separation value for a symmetric triplet at $\Omega=0, \beta=\pi/2$ is 
$\approx 3.35$. There is, of course, also a variety of asymmetric 
equilibria forming an infinite grid at $\Omega+\beta=\pi/2$. 

In multikink arrays the minimal separation between neighboring kinks 
tends to increase with the increasing size of the array. Early stages of 
evolution to equilibrium may appear to be rather disordered. An example of the relaxation 
process for a symmetric array of 11 kinks, starting from 
positions corresponding to the minimal stable separation for a 3-kink array, is shown in 
Fig.~2. When the number of kinks is large, there are alternative close lying equilibria even 
when separations are close to minimal one. Such equilibria were detected when the 
dynamic equations 
of a symmetric 15-kink array were integrated starting from slightly different initial 
conditions. Multikink arrays may form chaotic structures containing an arbitrary number of 
gaps of different size. 

At $\Omega+\beta$ differing from $\pm\pi/2$, the number of equilibria 
for kink-antikink pairs is finite, and is decreasing rapidly with growing 
deviation from the straight angle. The asymptotic estimate for a most 
distant equilibrium position is $x_{max} \asymp ( 2/\sqrt{\pi}) |\sin 
\beta/\cos(\Omega + \beta)|$. With no average bias $|P|^2-|U_0|^2$, 
equilibrium positions exist only while $\beta$ remains within the range 
of maximal variation of this curve; e.g. at $\Omega=0$ there are no 
solutions at $\beta>0.65 \pi$ or $\beta<0.41 \pi$.
 
In two dimensions, an additional factor is the line tension, that affects the 
kink motion through the curvature term in Eq.~(\ref{eqmot}) . An island 
of one state (say, $\chi \approx -1$) within a two-dimensional continuum 
of the alternative state would shrink, save for the optical interactions, 
when the phase equilibrium is unbiased. Due to nonlinear diffractional 
effects, the curved interphase boundary may stabilize at certain positions 
while retaining its circular shape because of the line tension. An {\em a 
priori} estimate for the radius $a$ of a stationary island is $a \approx 
\sigma/\alpha$. 

Solving Eq.~(\ref{basicdiff}) in polar coordinates with the initial condition 
\begin{equation}
E(x,0)=P \left[ e^{- i\beta} + 2i \sin \beta H(r-a) \right].
\label{inco2d}  \end{equation}
 yields the cavity field
\begin{eqnarray}
u(r)  &=&  e^{i\Omega} \left[ e^{-i\beta}+2 i \sin \beta \; v(r;a) \right];
\nonumber \\
  v(r,a) &=&  1 - a\int_0^\infty   e^{-i\lambda^2} 
       J_1(\lambda a)J_0(\lambda r) d\lambda ,
\label{vcyl}    \end{eqnarray}
where $J_n(x)$ is a Bessel function. Equilibrium radii $a$ verify the 
equation $a^{-1}= Q\,\mbox{Re} \, u(a)$, where $Q=2b\alpha|P|^2 
\sigma^{-1}$. At $r=a$, the integral in Eq.~(\ref{vcyl}) can be computed 
analytically:
 \begin{equation}
v(a;a)  =  1 +   J_0\left(\frac{a^2}{2}\right) \exp\left(\frac{ia^2}{2}\right) .
\label{vcyls}    \end{equation}
The resulting evolution equation of $a$ is
\begin{eqnarray}
 \frac{1}{\sigma}\,\frac{da}{dt} &=& \frac{Qh(a)\sin \beta- 1}{a} 
      - Q\cos \Omega \cos\beta;      \nonumber \\
h(a) &=& a\sin \left(\frac{a^2}{2}  + \Omega \right) 
      J_0 \left( \frac{a^2}{2} \right) .
\label{qecyl}  \end{eqnarray}
A convenient form of the equation defining equilibrium values of $a$ is
\begin{equation}
 (Q\sin \beta)^{-1} + a\cos \Omega \cot\beta =  h(a) .
\label{qcyl}  \end{equation}
Solutions of this equation can be obtained as intersections of a straight line 
presenting its left-hand side with the plot of the function $h(a)$ drawn in 
Fig.~3. The latter oscillates with an asymptotically constant amplitude 
between the maximum and minimum values $\frac{2}{\sqrt{\pi}}\cos ^2 
\left(\frac{\pi}{8} \right) \approx 0.963 $ and $-\frac{2}{\sqrt{\pi}}\sin  ^2 
\left(\frac{\pi}{8} \right) \approx -0.165 $. The number of solutions is 
infinite when either $\beta$ or $\Omega$ equals to $\pm \frac{\pi}{2}$, 
and the inverse of $Q\sin \beta$ lies within this interval. The stability 
condition with respect to radially symmetric perturbations is 
$ah'(a)<h(a)-1$. At $a \gg 1$, stable solutions lie on descending segments 
in Fig.~3, i.e. at $(2n-\frac{5}{4})\pi<a^2<(2n-\frac{1}{4})$. 

Stability against asymmetric perturbations dependent on the polar angle 
$\phi$ is tested \cite{pi94} by perturbing the interphase boundary,  $r 
=a[1+ \zeta (\phi,t)]$, while retaining the quasistationary approximation 
for the optical field. If $\zeta (\phi,t) \ll 1$, the curvature of the kink is 
expressed as $\kappa =a^{-1}(1- \zeta - \zeta_{\phi\phi})$. 
Expanding the long-range biasing field $w(r,\phi)$ in the vicinity of the 
unperturbed position brings Eq.(\ref{eqmot}) to the form
\begin{equation}
d\zeta/dt = a^{-2} \sigma\left( \zeta_{\phi\phi} + \zeta \right) +
          b \left[ \widetilde w(a,\phi) +  w'_s(a) \zeta \right] .
\label{smotd} \end{equation}
Here $w'_s(a)$ is the derivative of the stationary biasing field at the kink 
position obtained from Eq.~(\ref{vcyl}), and $\widetilde w(r,\phi)$ is the 
perturbation 
of the biasing field caused by the perturbation of the boundary. The latter 
field is obtained by solving Eq.~(\ref{basicdiff}) with the initial condition 
$E(r,\phi,0)=2 i\sin \beta \,\zeta (\phi) \sigma(a)$. Setting $\zeta (\phi,t) 
= \rho_n(t) \cos n\phi$ and adding, as usual, a constant phase shift, we 
obtain the scaled perturbation of the cavity field
\begin{eqnarray}
 \widetilde u (r,\phi;a) &=& 2i a \rho_n \sin \beta \,
           e^{i\Omega} \cos n\phi \;\widetilde u_n (r,;a); \nonumber \\
    \widetilde u_n (r;a) &=&\int_0^\infty \lambda  e^{-i\lambda^2} 
       J_n(\lambda a)J_n(\lambda r) d\lambda .
      \label{sVeqd}     \end{eqnarray}
At $n=1$, this expression is just the derivative $du/dx$ of the stationary 
field Eq.~(\ref{vcyl}) taken with the opposite sign; consequently, the two 
bracketed terms in Eq.~(\ref{smotd}) cancel, as they must do since this 
perturbation mode is equivalent to merely shifting the spot along the $x$ 
axis. Computing the integral in Eq.~(\ref{sVeqd}) to express the perturbed 
biasing field $\widetilde w(a,\phi) = 2 \alpha|P|^2 \mbox{Re} 
\,\widetilde u (r,\phi;a)$, and using it in  Eq.~(\ref{smotd}) yields the 
stability condition
\begin{equation}
 a^{-2} (n^2-1)>2 aQ \sin \beta \; [ h_n(a) - h_1(a)] ,
  \label{disp0d} \end{equation}
where
\begin{eqnarray}
 h_n(a)&=& \int_0^\infty \lambda  \sin (\lambda^2-\Omega) 
       J_n^2(\lambda a) d\lambda \nonumber \\
&=& \frac{1}{2} J_n\left( \frac{a^2}{2} \right) 
     \cos\left( \frac{a^2-n\pi}{2}-\Omega \right) .
      \label{sss}     \end{eqnarray}
At intermediate values of $a$, the quadrupole mode $n=2$, deforming 
the disk into an ellipse, causes instability on the ``stable'' descending 
branches in Fig.~3 below the inflection point. At very large values of $a$, 
the solution is unstable against either $n=3$ or $n=5$ mode, depending on 
the value of $\Omega$.
Instability may lead to the phenomenon of ``replicating spots'' similar to 
that observed in simulations of Turing patterns \cite{swin}. Spots may be 
arranged in crystalline arrays, though their shape remains nearly circular 
only at large separations.

{\em Acknowledgement.} This research has been supported by the 
Fund for Promotion of Research at the Technion.

%-------------------------------------------------------------------

\begin{figure}
\psfig{figure=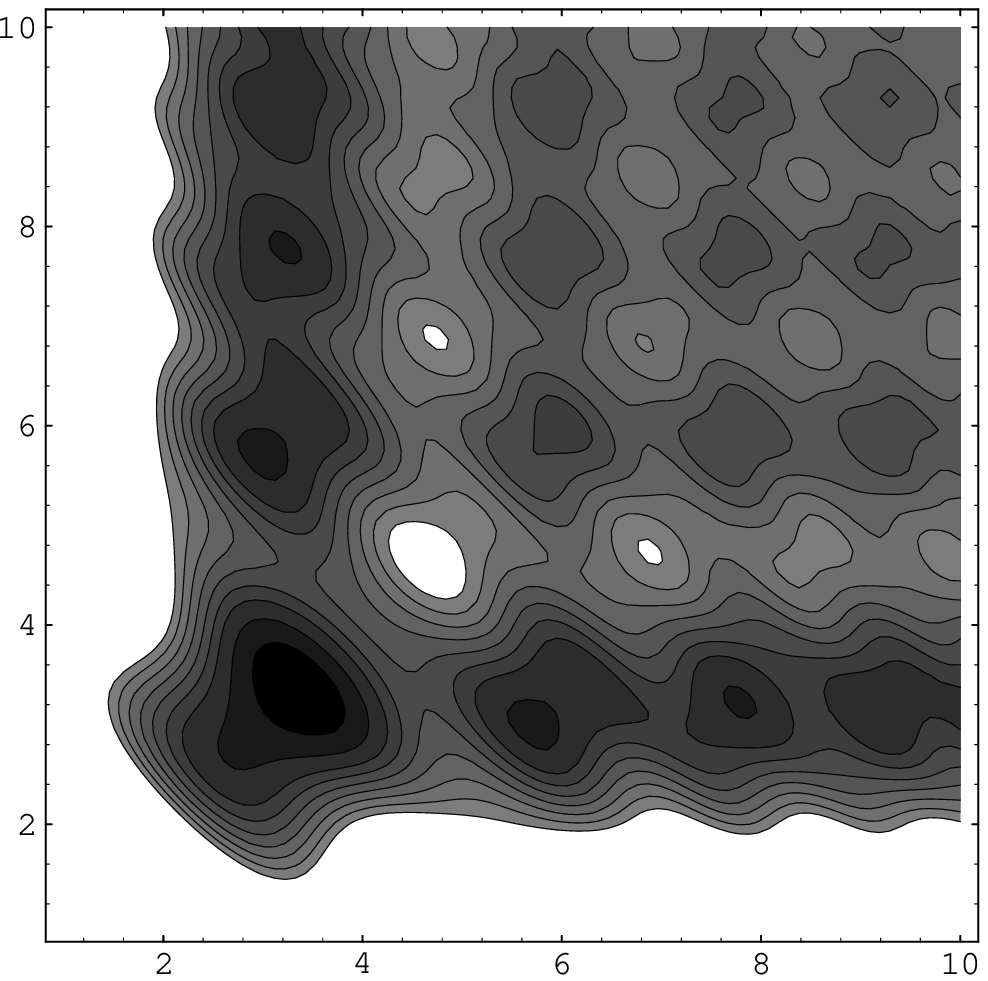}
\caption{The  relief of the potential for a kink -- antikink -- king triplet.}
\end{figure}

\begin{figure}
\psfig{figure=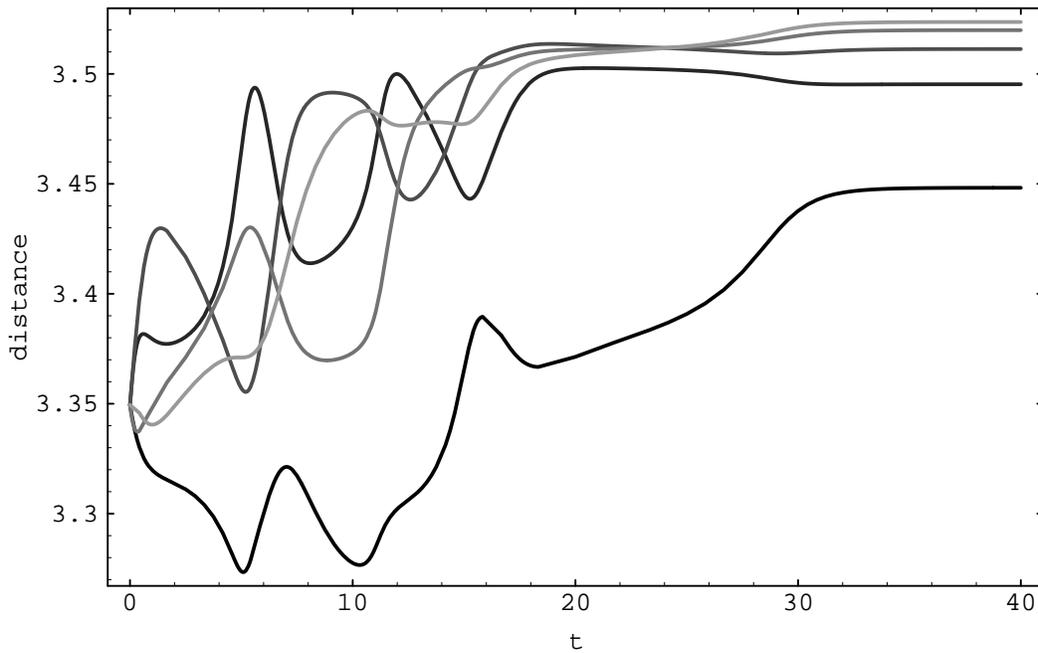}
\caption{Evolution of separation distances between kinks and antikinks 
in a symmetric 11-kink array ($\beta=\frac{\pi}{2},\Omega=0$). Darker 
curves correspond to kinks lying closer to the center.}
\end{figure}

\begin{figure}
\psfig{figure=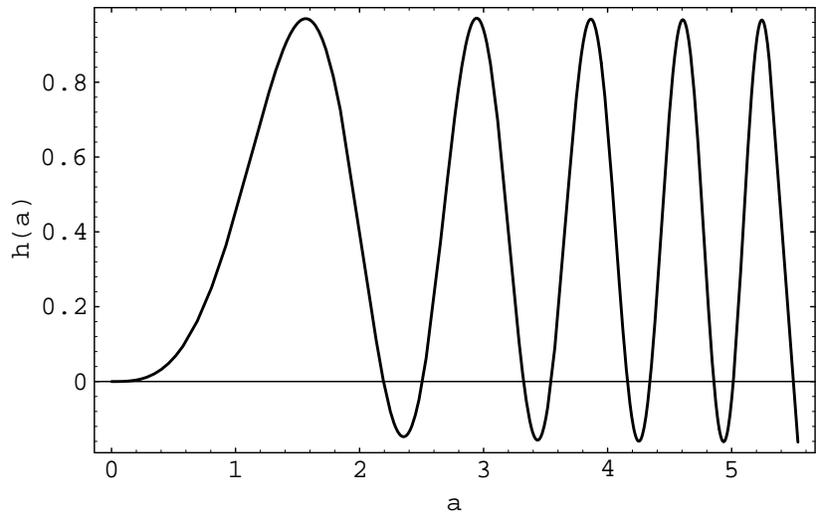}
\caption{The function $h(a)$ used for finding equilibrium radii of a 
solitary spot ($\Omega=0$).}
\end{figure}

\end{document}